\documentclass[a4paper,11pt]{article}
\pdfoutput=1 % if your are submitting a pdflatex (i.e. if you have
\usepackage{jheppub} % for details on the use of the package, please
\usepackage[T1]{fontenc} % if needed

\usepackage{graphicx}
\usepackage{dcolumn}
\usepackage{amsmath}
\usepackage{amssymb}
\usepackage{color}
\usepackage{hyperref}
\usepackage[utf8]{inputenc}
\usepackage[english]{babel}
\usepackage{dsfont}
\usepackage{xspace}
\usepackage{tabularx}
\usepackage{xcolor}
\usepackage{csquotes}
\usepackage{mathtools}
\usepackage{physics}
\usepackage{comment}

\newcommand{\tfd}{\text{TFD}}
\renewcommand{\i}{\text{i}}

\newcommand{\1}{\mathds{1}}

\title{\boldmath Geometric Quantum Discord Signals Non-Factorization}

\author{Souvik Banerjee, Pablo Basteiro, Rathindra Nath Das and Moritz Dorband}

\affiliation{Institute for Theoretical Physics and Astrophysics and Würzburg-Dresden Cluster of Excellence ct.qmat, Julius-Maximilians-Universität Würzburg, Am Hubland, 97074 Würzburg, Germany}

\emailAdd{\{souvik.banerjee, pablo.basteiro, das.rathindranath, moritz.dorband\}@uni-wuerzburg.de}

\abstract{We propose the information-theoretic quantity of geometric quantum discord (GQD) as an indicator of the factorization properties of a given quantum system. In particular, we show how non-vanishing discord implies that the corresponding partition function does not factorize, both for generic pure states and the thermofield double state as a state with a known geometric dual in light of the AdS/CFT correspondence. Via this analysis, we give a novel interpretation to the \textit{thermomixed double} state as the best purely classical approximation of the Einstein--Rosen bridge. We connect the non-vanishing of GQD with the existence of wormhole microstates.}

\begin{document} 
\maketitle
\flushbottom

\section{\label{sec:Introduction}Introduction}

The AdS/CFT correspondence \cite{Maldacena:1997re,Witten:1998qj,Gubser:1998bc} has allowed for a plethora of progress in high-energy physics and has provided us with new insights into a seemingly fundamental connection between gravity and quantum field theories, known as the holographic principle \cite{tHooft:1993dmi,Susskind:1994vu,Susskind:1998dq}. Quantum information measures, such as entanglement entropy \cite{Ryu:2006bv,Ryu:2006ef,Hubeny:2007xt} and circuit complexity \cite{Susskind:2014rva,Stanford:2014jda,Susskind:2014moa} have played a paramount role in developing and augmenting the holographic dictionary.

Nevertheless, certain instances arise where the holographic correspondence between the bulk and boundary Hilbert spaces leads to apparent contradictions. For example, the so-called \emph{thermofield double} (TFD) state, a weighted superposition of two copies of CFT energy eigenstates, is dual to an eternal two-sided black hole in an AdS spacetime \cite{Maldacena:2001kr}. This state can be thought of as a purification of the thermal state given by the exterior of a black hole at a finite Hawking temperature. The Hilbert space associated to the whole boundary quantum field theory is assumed to factorize, i.e. $\mathcal{H}_{\tfd}=\mathcal{H}_L\otimes\mathcal{H}_R$, since the two copies of the CFT do not share an explicit interaction. In contrast, the eternal black hole represents a smooth geometry connecting the left and the right boundaries through a non-traversable wormhole, realized as the Einstein--Rosen bridge. From the holographic perspective, the Einstein--Rosen bridge is a geometric manifestation of the entanglement between the dual CFTs living on the left and the right boundaries giving rise to the maximally entangled TFD state. This surprising connection between a classical geometry and the quantum entanglement has been coined as the ER $=$ EPR (Einstein--Rosen $=$ Einstein--Podolski--Rosen) \cite{VanRaamsdonk:2010pw,Maldacena:2013xja, Verlinde:2022dtx} proposal. The existence of the smooth bulk wormhole geometry implies that the bulk Hilbert space cannot assume a factorized form like its dual boundary counterpart. This apparent ambiguity between the basic structures of the bulk and the boundary Hilbert spaces gives rise to the so-called \textit{factorization puzzle}. 

It is apparent that the black hole introduces correlations between both sides of the TFD state. Indeed, the two CFTs in the TFD state are maximally entangled with each other. Nevertheless, the TFD is a very fine-tuned pure state with zero von Neumann entropy. In order to match the fact that a realistic black hole is a macroscopic object with enormous entropy, consistent with the size of its microscopic phase space, one would naturally expect the state of the black hole to actually be represented by a mixed state, instead of a pure state. In particular, this mixed state should be decohered in such a way that the correlations measured by an asymptotic observer remain unchanged. Motivated by this, it was recently argued in \cite{Verlinde:2020upt} that from the perspective of a low-energy one-sided observer, the typical state of two-sided black hole should rather be described by the so-called \textit{thermomixed double} (TMD) state, which contains purely classical correlations and does not support any quantum correlations.
Such a state would still imply the non-factorization of the bulk Hilbert space, but it would not comply with the EPR side of the conjecture, since no quantum entanglement is present. Rather, the two sides, in this case, only share the classical Shannon mutual information. This state of affairs suggests that quantum correlations not encompassed by entanglement entropy play an important role in the understanding of the factorization puzzle.

It is therefore important to study different measures of entanglement which capture quantum correlations beyond the entanglement entropy, one example of such a measure being quantum discord \cite{Zurek2000Einselection,Ollivier2001Quantum,henderson2001classical}. For the particular example of the isotropic state, which is a one-parameter family of states interpolating between pure Bell states and fully mixed states, it is known that this measure of quantum correlation is non-zero even when other measures of entanglement, such as entanglement of formation and entanglement negativity, vanish, see e.g. \cite{Ollivier2001Quantum,Dajka_2012} (and also \cite{Poxleitner2021Gaussian} for an explicit study of quantum discord for a Gaussian isotropic state). Unfortunately, due to its definition involving a minimization over projective measurements, quantum discord is generally hard to compute. In fact, it has been shown that computing quantum discord is NP complete \cite{Huang2014NPcomplete}. Therefore, as an alternative to quantum discord, the measure known as \emph{geometric quantum discord} (GQD) \cite{Daki2010necessary,Bera2017Quantum} was established. It was shown in \cite{Daki2010necessary} that a non-vanishing GQD implies the non-vanishing of quantum discord. GQD also provides an upper bound on entanglement negativity \cite{PhysRevA.84.052110}.

In this work, we propose the GQD as an easily computable indicator of non-factorization in the boundary. We show how this measure signals the existence of purely quantum correlations in generic pure states. This manifests as non-factorization of the modular partition function of the corresponding system. We find that the GQD for pure states can always be expressed by the second Rényi entropy of a particular classical state. As a byproduct of this, we find that for pure states the GQD is equivalent to the \enquote{geometric measure of quantum correlation} first introduced in \cite{Guo:2020rwj}. Furthermore, utilizing the TFD state as a prime example of a holographic state, we show that non-zero GQD implies non-factorization of the boundary Hilbert space as indicated by non-factorization of the thermal partition function. Our analysis enables for a novel interpretation of the TMD state discussed in \cite{Verlinde:2020upt} as the \emph{best classical approximation to the TFD state}. Moreover, we argue that our reasoning applies in principle to other black hole states as well, e.g. states dual to black holes formed by collapse. Some comments on obtaining the GQD purely by a bulk computation are also provided. Finally, we compare our proposed diagnostic for non-factorization with the non-exactness criterion previously suggested in \cite{Verlinde:2021kgt}.

\section{\label{sec:QD}Brief Review of Quantum Discord}

The fact that quantum discord quantifies purely quantum correlations not captured by entanglement entropy is best understood by looking at its definition in terms of more familiar quantum information measures. We thus provide a brief introduction to quantum discord and its properties in the following.

Introduced independently in \cite{Zurek2000Einselection,Ollivier2001Quantum} and \cite{henderson2001classical}, quantum discord is defined as the difference between two classically equivalent ways of writing the mutual information. Let $A$ and $B$ be two random variables. Classically, the mutual information can be defined as,
\begin{align}
	I(A:B)=H(A)+H(B)-H(A,B)\,,\label{eq:MutualInformation}
\end{align}
or, equivalently,
\begin{align}
	I(A:B)=H(A)-H(A|B)\,,\label{eq:MutualInformation2}
\end{align}
where $H$ denotes the classical Shannon entropy and $H(A|B)=H(A,B)-H(B)$ is the conditional entropy. For a quantum system described by a total density matrix $\rho_{AB}$, the random variables can be replaced by the reduced density matrices of the individual subsystems $\rho_A=\Tr_B(\rho_{AB})$ (and the same with $A\leftrightarrow B$). Shannon entropies $H$ are then replaced by von Neumann entropies $S$. This recipe, though not problematic for the definition in \eqref{eq:MutualInformation}, does not lead to a positive-definite expression for definition \eqref{eq:MutualInformation2}, due to the presence of the conditional entropy $H(A|B)$ \cite{Bera2017Quantum}. The interpretation of conditional entropy for quantum systems is not clear a priori, since it involves knowledge of one of the subsystems, implying some kind of measurement has been performed. A possible definition was proposed in \cite{Ollivier2001Quantum} and is given by
\begin{align}
	S(A|B)=\min_{\Pi_k^B\in{\cal M}^B}\sum_kq_kS(\rho_k^A)\,,\label{eq:QuantumConditionalEntropy}
\end{align}
where $\Pi_k^B$ denotes a projective measurement performed on $B$, whose outcome is labeled by $k$. The set of all such measurements is denoted by ${\cal M}^B$. The reduced density matrix
\begin{align}
	\rho_k^A&=\frac{1}{q_k}\Tr_B\left[\big(\mathds{1}_2\otimes\Pi_k^B\big)\rho_{AB}\big(\mathds{1}_2\otimes\Pi_k^{B\,\dagger}\big)\right]\label{eq:ReducedDensityMatrix}\,,\\
	\text{where}\quad q_k&=\Tr\left[\big(\mathds{1}_2\otimes\Pi_k^B\big)\rho_{AB}\big(\mathds{1}_2\otimes\Pi_k^{B\,\dagger}\big)\right]\label{eq:ProbabilityOfReducedDensityMatrix}\,,
\end{align}
describes the subsystem $A$ provided that outcome $k$, with probability $q_k$, was measured in $B$. Using \eqref{eq:QuantumConditionalEntropy}, the mutual information in \eqref{eq:MutualInformation2} is then written as
\begin{align}
	J(A:B)=S(\rho_A)-S(A|B)\,.
\end{align}
This quantity captures classical correlations only \cite{henderson2001classical} and is known as asymmetric mutual information due to the presence of the second term, which spoils the symmetry under the exchange $A\leftrightarrow B$. Quantum discord is then defined as
\begin{align}
	Q(A:B)=I(A:B)-J(A:B)\,,\label{eq:QuantumDiscord}
\end{align}
and does not vanish in general. Importantly, since the difference between \eqref{eq:MutualInformation} and \eqref{eq:MutualInformation2} always vanishes for classical setups, quantum discord describes correlations between the systems $A$ and $B$ which are guaranteed to be purely of quantum nature. In particular, quantum discord can be non-zero even though the entropy of formation vanishes \cite{Ollivier2001Quantum}. This means that quantum discord is sensitive to quantum correlations that can not be detected by the entropy of formation. Moreover, \eqref{eq:QuantumDiscord} reduces to the entanglement entropy for a pure state $\rho_{AB}=|\psi\rangle\langle\psi|$, since in this case $S(\rho_{AB})=0=S(A|B)$. Finally, $Q(A:B)\geq0$, vanishing if and only if the total correlation, quantified by $I$, is purely classical, such that it is canceled exactly by $J$.

\section{\label{sec:GQD}Geometric Quantum Discord for Pure States}

Unfortunately, explicit computations of quantum discord are scarce and in general only accessible numerically, due to the complicated minimization procedure over projective measurements involved in its definition, cf.~\eqref{eq:QuantumConditionalEntropy}. Moreover, it is unclear how to define quantum discord in quantum field theories, since the notion of projective measurements in this setting is not evident in general (although see \cite{Antonini:2022lmg,Antonini:2023aza} for recent work in this direction). As a partial resolution to this issue, the authors of \cite{Daki2010necessary} introduced the GQD (see \cite{Bera2017Quantum} for a review) whose computation does not involve a minimization over measurements, but rather a more feasible minimization over states,
\begin{align}
	Q^{(2)}(A:B)=\min_{\chi\in\Omega_0}||\rho_{AB}-\chi||^2\,.\label{eq:GeometricQuantumDiscord}
\end{align}
Here, $\rho_{AB}$ is the density matrix of the full quantum system, $||X||^2=\Tr(X^2)$ is the Hilbert-Schmidt norm, and $\Omega_0$ is the set of zero-discord states. These zero-discord states are given by quantum-classical (q-c) states
\begin{align}
	\chi^{\text{q-c}}=\sum_kq_k\rho^A_k\otimes|\phi_k^B\rangle\langle\phi_k^B|\,,\label{eq:QuantumClassicalState}
\end{align}
where $q_k\geq0$, $\sum_k q_k=1$ and $\langle\phi^B_k|\phi^B_l\rangle=\delta_{kl}$. The density matrix $\rho^A_k$ is the quantum state describing the subsystem $A$.\footnote{If one chooses to define the measure with respect to subsystem $A$ instead of $B$, the definition of zero-discord states is analogously obtained by exchanging $\rho_k\leftrightarrow|\phi_k\rangle\langle\phi_k|$. These are then denoted as classical-quantum (c-q) states.} It was proven in \cite{Daki2010necessary} that the non-vanishing of the GQD provides a necessary and sufficient condition for the non-vanishing of the quantum discord \eqref{eq:QuantumDiscord}. Due to the definition of the GQD in terms of the norm, the state minimizing $Q^{(2)}(A:B)$ is interpreted as the q-c state that best approximates the full state $\rho_{AB}$.

Owing to the ubiquitous role of pure states in quantum field theories and holography, in the remainder of this paper, we will evaluate $Q^{(2)}(A:B)$ for pure states $\rho_{AB}=|\psi\rangle\langle\psi|$. Any pure state $|\psi\rangle$ in a bipartite Hilbert space ${\cal H}_{AB}$ can be written in Schmidt decomposition
\begin{align}
    |\psi\rangle=\sum_k\lambda_k|\phi_k^A\rangle|\phi_k^B\rangle\,,\label{eq:SchmidDecomposition}
\end{align}
such that
\begin{align}
    \rho_{AB}=\sum_{k,l}\lambda_k\lambda_l|\phi_k^A\rangle|\phi_k^B\rangle\langle\phi_l^A|\langle\phi_l^B|\,,\label{eq:PureStateDensityMatrix}
\end{align}
where $\lambda_k$ are the Schmidt coefficients which determine the entanglement properties of $|\psi\rangle$. In particular, $\lambda_k^2$ are the eigenvalues of the reduced density matrix $\rho_{B/A}=\Tr_{A/B}(\rho_{AB})$. Computing the Hilbert--Schmidt norm of the difference of $\rho_{AB}$ \eqref{eq:PureStateDensityMatrix} and a generic q-c state \eqref{eq:QuantumClassicalState} yields
\begin{align}
    \norm{\rho_{AB}-\chi^{\text{q-c}}}=1-2\sum_k\lambda_k^2q_k\Tr_A(\rho_k^A)+\sum_kq_k^2\Tr_A(\rho_k^A\rho_k^A)\,.\label{eq:NormExpression}
\end{align}
While $\Tr_A(\rho_k^A)=1$ by definition, the trace in the third term requires more care. The trace of the square of a density matrix is generally known as its purity, so for any $\rho$, $\Tr(\rho^2)=1$ if and only if $\rho$ is pure. If $0<\Tr(\rho^2)<1$, $\rho$ is mixed. To find the minimum of \eqref{eq:NormExpression}, we compute its derivative with respect to $q_k$ and set it to zero, resulting in
\begin{align}
    q_k=\frac{\lambda_k^2}{\Tr(\rho_k^A\rho_k^A)}.\label{eq:QkMinimization}
\end{align}
However, if the purity of $\rho_k^A$ does not equal to one, the expression in \eqref{eq:NormExpression} can become negative after inserting \eqref{eq:QkMinimization}, which has to be excluded due to the definition of the norm. This is consistent with computing \eqref{eq:ReducedDensityMatrix} and \eqref{eq:ProbabilityOfReducedDensityMatrix} explicitly for \eqref{eq:SchmidDecomposition}, resulting in
\begin{align}
    q_k=\lambda_k^2\quad\text{and}\quad\rho_k^A=|\phi_k^A\rangle\langle\phi_k^A|\,.\label{eq:QkDirectly}
\end{align}
Note that this is not mere coincidence. By definition, for a pure state $Q(A:B)=S(\rho^A)$. Operationally, this is linked to the fact that $S(A|B)=0$ for a pure state $\rho_{AB}$. Since $S(A|B)$ is defined as a sum over individual entropies $S(\rho_k^A)$ weighed by $q_k$, this vanishes exactly when each $\rho_k^A$ is given by a pure state, as found in \eqref{eq:QkDirectly} and by the aforementioned argument on the purity of $\rho_k^A$.

Inserting the above results for $q_k$ and the purity of $\rho_k^A$ into \eqref{eq:NormExpression}, we find that GQD for a generic pure state \eqref{eq:SchmidDecomposition} is given by
\begin{align}
    Q^{(2)}(A:B)=1-\sum_k\lambda_k^4\,.\label{eq:GQDPureState}
\end{align}
Since $\rho_{AB}$ is pure, we know that for the quantum discord holds $Q(A:B)=S(\rho_A)$, which is non-vanishing in general. Since $\sum_k\lambda_k^2=1$ with $0\leq\lambda_k^2\leq1$, the result for $Q^{(2)}(A:B)$ given in \eqref{eq:GQDPureState} vanishes if and only if some $\lambda_{k^\ast}=1$ with all other $\lambda_{k\neq k^\ast}$ vanishing. Thinking back to the Schmidt decomposition \eqref{eq:SchmidDecomposition}, this configuration of $\lambda_k$'s is precisely the case where $|\psi\rangle$ is a product state and $Q(A:B)=S(\rho_A)=0$.

Our above results have a further interesting consequence, namely they show that the q-c state which minimizes the Hilbert--Schmidt norm is
\begin{align}
    \chi_0^{\text{q-c}}=\sum_k\lambda_k^2|\phi_k^A\rangle\langle\phi_k^A|\otimes|\phi_k^B\rangle\langle\phi_k^B|\,\widehat{=}\, \chi_0^{\text{c-c}}\,.\label{eq:q-cStateMinimizingGQD}
\end{align}
which is actually a classical-classical (c-c) state. In other words, the minimization in \eqref{eq:GeometricQuantumDiscord}, which is a priori over the set $\Omega_0$ of all zero-discord states, actually constraints $\chi^{\text{q-c}}$ to be a c-c state. This connects nicely to the measure $\bar{Q}^{(2)}(A:B)$ introduced in \cite{Guo:2020rwj} and denoted as \enquote{geometric measure of quantum correlations},
\begin{align}
    \bar{Q}^{(2)}(A:B)=\min_{\chi\in\Lambda_0}\frac{\norm{\rho_{AB}-\chi}^2}{\norm{\rho_{AB}}^2}\,,
\end{align}
which is analogous to \eqref{eq:GeometricQuantumDiscord} except that the minimization is carried out over the set $\Lambda_0$ of c-c states. The latter are of the form
\begin{align}
    \chi^{\text{c-c}}=\sum_{k,l}q_{kl}|\phi_k^A\rangle\langle\phi_k^A|\otimes|\phi_l^B\rangle\langle\phi_l^B|\,.
\end{align}
Thus, even though $\Lambda_0\subset\Omega_0$, given a pure state $\rho_{AB}$ as starting point the expressions for $Q^{(2)}(A:B)$ and $\bar{Q}^{(2)}(A:B)$ can be considered equivalent. While the sets $\Omega_0$ and $\Lambda_0$ are of course different, the state minimizing $Q^{(2)}(A:B)$ is the same state minimizing $\bar{Q}^{(2)}(A:B)$. Considering mixed states however, the two measures will be different in general. For instance, for the isotropic state of a two spin system
\begin{align}
    \rho_{\text{iso}}=\frac{1-z}{4}\1_4+\frac{z}{2}\big(|00\rangle+|11\rangle\big)\big(\langle00|+\langle11|\big),\label{eq:IsotropicState}
\end{align}
the GQD follows as
\begin{align}
    Q_{\text{iso}}^{(2)}(A:B)=\frac{z^2}{2}.
\end{align}
The minimization dictates that \eqref{eq:QuantumClassicalState} consists of
\begin{align}
    q_{0/1}^{\text{iso}}=\frac{1}{2},\quad\rho_0^{A,\,\text{iso}}&=\begin{bmatrix}\frac{1+z}{2}&0\\0&\frac{1-z}{2}\end{bmatrix},\quad\rho_0^{B,\,\text{iso}}=\begin{bmatrix}\frac{1-z}{2}&0\\0&\frac{1+z}{2}\end{bmatrix}.
\end{align}
Since the $\rho_i^A$ are not pure for general $z$, $\chi_0^{\text{q-c}}$ does not reduce to a c-c state and therefore cannot be the state minimizing $\bar{Q}^{(2)}(A:B)$. While the $\rho_i^A$ are pure for $z=1$, in this case also the isotropic state \eqref{eq:IsotropicState} itself reduces to a pure state.

Finally, we supplement the above analysis by pointing out that $Q^{(2)}(A:B)$ in \eqref{eq:GQDPureState} can be expressed in terms of the second Rényi entropy
\begin{align}
    S^{(k)}(\rho)&=\frac{1}{1-k}\ln\Tr\big(\rho^k\big)\quad\text{for}\quad k=2\,,\\
    \text{resulting in}\quad&Q^{(2)}(A:B)=1-e^{-S^{(2)}(\chi_0^{\text{q-c}})}\,.\label{eq:GQDRenyi}
\end{align}
Introducing the modular Hamiltonian $H_{\text{mod}}\equiv-\ln\chi_0^{\text{q-c}}$, we also write
\begin{align}
    Q^{(2)}(A:B)=1-\frac{Z(2H_{\text{mod}})}{Z^2(H_{\text{mod}})}\,,
    \label{eq:GQDModularHamiltonian}
\end{align}
where $Z(H_{\text{mod}})$ is the partition function with $\beta H$ replaced by the modular Hamiltonian. Therefore, our analysis shows that there are non-vanishing quantum correlations between the subsystems $A$ and $B$ if the modular partition function does not factorize. In other words, the Schmidt basis makes the non-factorizing nature of a quantum system manifest in terms of the non-factorization of the modular partition function. Moreover, in view of the replica trick \cite{Holzhey:1994we,Calabrese:2004eu}, this rewriting makes explicit that the computation of the GQD is accessible in field theory.

\section{\label{sec:QDTFD}Geometric Quantum Discord for the Thermofield Double State}

We now turn to a holographic setting and analyze the GQD for the TFD state. In particular, we provide a unifying analysis which puts our above computations under the scope of the wormhole setup of \cite{Verlinde:2020upt}. We briefly introduce the TFD and TMD states and show how evaluating the GQD \eqref{eq:GeometricQuantumDiscord} for the TFD state provides a clear diagnostic of non-factorization from the boundary perspective. Furthermore, our analysis yields a novel interpretation of the TMD state.

Given two copies of a CFT, the TFD state is defined as
\begin{align}
	|\text{TFD}\rangle=\frac{1}{\sqrt{Z(\beta)}}\sum_ne^{-\beta\frac{E_n}{2}}|n_L\rangle|n_R\rangle^\ast\,,
\label{eq:TFD}
\end{align}
where $Z(\beta)=\sum_ne^{-\beta E_n}$ is the thermal partition function and $\beta$ is the inverse temperature. The energy eigenstates $|n_i\rangle$ have energy eigenvalues $E_{n,i}$, where $E_{n,L}=E_{n,R}$ and we denote $E_n=\frac{1}{2}(E_{n,L}+E_{n,R})$. The state $|n_R\rangle^\ast$ denotes the conjugation of $|n_R\rangle$ by an anti-unitary operator, $|n_R\rangle^\ast=\Theta|n_R\rangle$, where $\Theta$ typically represents $\cal CPT$.\footnote{$\cal CPT$ is the common choice, especially in the context of field theory. More general, any anti-unitary operator can be written as a unitary operator $V$ times complex conjugation $K$. Complex conjugation on the other hand is related to the time reversal operator, $K\sim\cal T$. Therefore, the anti-unitary operator $\Theta$ can always be written as $V\cal T$ with $V=\cal CP$, but any other choice for $V$ is equally valid, albeit potentially less convenient for the application to physics.} In the following, we will denote $p_n=\frac{1}{\sqrt{Z(\beta)}}e^{-\beta\frac{E_n}{2}}$ for convenience. These $p_n$ are the Schmidt coefficients of the TFD state.

The TFD state is a pure state description of the two-sided eternal black hole. However, it is argued in \cite{Verlinde:2020upt} that right after the black hole is put in contact with its ambient spacetime, interaction leads to decoherence and the system evolves into a mixed state. This process turns quantum into classical correlations. The form of this mixed state is constrained by the fact that for the one-sided low-energy observer, the state of the black hole looks thermal both before and after decoherence. The way in which the quantum information in the TFD state is affected by the decoherence can be understood as follows. Given \eqref{eq:TFD}, a natural way to store quantum information within the state is to include phase factors $\alpha_n$,
\begin{align}
	|\text{TFD}\rangle_\alpha=\sum_np_ne^{\text{i}\alpha_n}|n_L\rangle|n_R\rangle^\ast\,.\label{eq:TFDwithPhases}
\end{align}
These phases can be understood as arising from time evolution of the TFD state by the operator $H_L+H_R$, $H_{L/R}$ being the boundary Hamiltonians. Since the CFT spectrum is highly random, for any instance of time $t$ there is a phase factor $e^{\text{i}\alpha_n}$ for the corresponding energy $E_n$. The time coordinate $t$ used for this time evolution is not defined globally due to the jump of the timelike Killing vector at the black hole horizon. This jump in time is however invisible to any local observer \cite{Papadodimas:2015jra}. Therefore, the phases $\alpha_n$ can be understood as storing quantum information \cite{Verlinde:2020upt}. As far as the local low-energy observer is concerned, the states \eqref{eq:TFDwithPhases} describe the same classical geometry for any set of phases $\{\alpha\}$. Therefore, from her perspective, it is reasonable to consider an incoherent sum over all of those states instead of one particular instance. This leads to the definition of the TMD state \cite{Verlinde:2020upt},
\begin{align}
	\rho_{\text{TMD}}=\frac{1}{N}\sum_{\{\alpha\}}|\text{TFD}\rangle_\alpha\,_\alpha\langle\text{TFD}|\,.\label{eq:TMD}
\end{align}
Due to the highly random nature of the typical CFT spectrum, the sum can be performed using \mbox{$\frac{1}{N}\sum_{\{\alpha\}}e^{\text{i}(\alpha_n-\alpha_m)}=\delta_{nm}$}, resulting in
\begin{align}
	\rho_{\text{TMD}}=\sum_np_n^2|n_L\rangle\langle n_L|\otimes|n_R\rangle\langle n_R|\,.
\end{align}
This state describes a fully decohered two-sided black hole after interactions with the environment, i.e. all quantum entanglement has been converted into classical correlation.

We are now in position to analyze the GQD for the TFD state. Inserting the Schmidt coefficients $p_n$ of the TFD state into our above result \eqref{eq:GQDPureState} yields
\begin{align}
    Q^{(2)}(L:R)=1-\sum_np_n^4\,.\label{eq:GQD-TFD-Result}
\end{align}
We therefore find that $Q^{(2)}(L:R)$ indicates the existence of quantum correlations in one particular instance of the TFD state.\footnote{Note that the computation of \eqref{eq:GeometricMeasureForTFD} could also be performed with $|\text{TFD}\rangle_\alpha$ instead of $|\text{TFD}\rangle$, without changing the result. The additional phases drop out due to the definition of the norm.} This statement itself is of course not a surprise: it is well known that the TFD state carries quantum correlations between the left and right CFT in the sense of EPR pairs, measured by the entanglement entropy. What is however remarkable is the specific state singled out by the minimization involved in computing the GQD for the TFD state. Following our above analysis of generic pure states, we know that this is a c-c state. Given our earlier result \eqref{eq:q-cStateMinimizingGQD}, the c-c state minimizing the GQD for the TFD state is precisely the TMD state,
\begin{align}
    \chi_0^{\text{c-c}}=\sum_np_n^2|n_L\rangle\langle n_L|\otimes|n_R\rangle\langle n_R|=\rho_{\text{TMD}}\,. 
\end{align}
We emphasize that this derivation of the TMD state is purely quantum information theoretic. In particular, it does not rely on considerations about the dual gravitational system, such as interactions between the black hole and its ambient spacetime as in \cite{Verlinde:2020upt}. Still, our derivation allows for a novel yet fitting interpretation of the TMD state: it arises as \emph{the state that best approximates the TFD state without including quantum correlations between the EPR pairs}. In terms of the dual geometry, there are only classical correlations across the horizon.

Inserting the explicit values for $p_n$ enables us to write the GQD in terms of the thermal partition function. The resulting expression is reminiscent of that obtained from Euclidean path integral considerations in \cite{Verlinde:2020upt,Verlinde:2021kgt,Verlinde:2021jwu},
\begin{align}
    Q^{(2)}(L:R)=1-\frac{\sum_ne^{-2\beta E_n}}{Z^2(\beta)}=1-\frac{Z(2\beta)}{Z^2(\beta)}\,.\label{eq:GeometricMeasureForTFD}
\end{align}
The fact that the GQD for the TFD state can be written in this way is not a mere coincidence: as shown in \cite{Verlinde:2020upt,Verlinde:2021kgt}, the TMD state can be used to define the gravitational partition function of $k$-fold wormholes by
\begin{align}
    \Tr\big(\rho_{\text{TMD}}^k\big)=\frac{Z(k\beta)}{Z^k(\beta)}\,.\label{eq:TrTMDPartitionFunction}
\end{align}
Combining this with the relation between the GQD and the second Rényi entropy \eqref{eq:GQDRenyi}, the result \eqref{eq:GeometricMeasureForTFD} follows. We will discuss in Sec.~\ref{sec:Microstates} how \eqref{eq:GeometricMeasureForTFD} arises from self-averaging of black hole microstates. Note that, as explained above in \eqref{eq:GQDModularHamiltonian}, the relation between the GQD and the thermal partition function is not true for a generic pure state, but has to be adapted to the modular partition function. This happens because the entanglement spectrum of the TFD state is thermal, while the entanglement spectrum of an arbitrary pure state will be more generic, so the modular partition function of \eqref{eq:GQDModularHamiltonian} becomes the thermal one for the TFD state.

The result \eqref{eq:GeometricMeasureForTFD} exhibits several interesting features. First, notice that \eqref{eq:GeometricMeasureForTFD} does not vanish in general. According to the necessary and sufficient condition derived in \cite{Daki2010necessary}, this means that also the quantum discord itself does not vanish. This is in accordance with the fact that the TFD state is pure and hence $Q(L:R)=S(\rho_L)$. Since the reduced density matrix of the TFD state is the thermal mixed state, $S(\rho_L)\neq0$. Second, we find that $Q^{(2)}(L:R)$ vanishes if and only if $Z(2\beta)=Z^2(\beta)$, implying factorization. As a consistency check, this equality is true for $\beta\to\infty$. In this limit, the TFD state reduces to a product state $\lim_{\beta\to\infty}|\tfd\rangle=|0_L\rangle|0_R\rangle$, so there are no EPR pairs shared by the left and right CFT. From the bulk perspective, this limit is understood as removing the black hole interior, i.e. the region responsible for the wormhole interpretation. This implies factorization also in the bulk since there are no quantum correlations to support the ER bridge.

From the analysis above, our main result can be summarized as follows. A non-vanishing discord implies the existence of quantum correlations. Using the analysis of the TMD state performed in \cite{Verlinde:2020upt}, we are able to give the condition for non-vanishing discord \eqref{eq:GeometricQuantumDiscord} introduced in \cite{Daki2010necessary} an interpretation in terms of wormhole physics. For pure states, as we have shown in \eqref{eq:GQDModularHamiltonian} and \eqref{eq:GeometricMeasureForTFD}, a non-vanishing discord is present if and only if the modular or thermal partition function does not factorize. In the limit $\beta\to\infty$ of the TFD state, which effectively puts the bulk at temperature $T=0$ and removes the wormhole, both $Q^{(2)}(L:R)$ and the discord $Q(L:R)=S(\rho_L)$ vanish. We have shown that the relation \eqref{eq:GQDRenyi} holds for generic pure states and thus we argue that there is a general correspondence,
\begin{equation}
    \begin{split}
        \textit{non-zero}\,&\,\textit{quantum discord}\\
        &\Downarrow\\
        \textit{non-factorization due}\,&\,\textit{to quantum correlations}\,.\notag
    \end{split}
\end{equation}
Note that non-factorization may also result from classical correlations. In this case however, the off-diagonal contributions in $\rho_{\tfd_\alpha}$, responsible for the ``quantumness'', go to zero and $\rho_{\tfd_\alpha}\to\rho_{\text{TMD}}$ as a state. 

We can therefore formulate the main statement of our results in the following way. Given a quantum state $\rho_{AB}$ describing the entirety of a system, we analyze whether its corresponding (modular) partition function factorizes. Although it might be possible to prove this by explicit computation for finite-dimensional systems, it becomes non-trivial in usual holographic setups. This is because there are typically no a priori arguments from the boundary perspective as to why the partition function should not factorize. Our results offer a novel, practical way of answering this question via the computation of geometric quantum discord. By finding the closest classical approximation $\rho^{c-c}_{AB}$ to the original quantum state $\rho_{AB}$, we can straightforwardly insert it into \eqref{eq:GQDRenyi} and see if its modular partition function factorizes. Finally, if the partition function of $\rho^{c-c}_{AB}$ does not factorize, then the partition function of $\rho_{AB}$ does not factorize either, since both states share the same Schmidt coefficients. Thus, the diagnosis of non-factorization for generic pure quantum states, and therefore the presence of wormhole-like physics, reduces to the computation of second Rényi entropies for their closest classical approximations\footnote{Factorization can be assessed through classical correlations, but computing GQD avoids the direct computation of the partition function. The minimization procedure emulates classical correlations within a quantum state. In other words, GQD makes the role of entanglement in non-factorization manifest by selecting purely quantum correlations present in a given state.}.

From the holographic perspective, the quantum correlation indicated by the GQD can be understood as follows. The TFD state, as well as its generalized version \eqref{eq:TFDwithPhases} and the TMD state all look the same to the local low-energy observer. Correspondingly, also the two-point function is the same irrespective of whether it is evaluated on $|\tfd\rangle$, $|\tfd\rangle_\alpha$ or $\rho_{\text{TMD}}$ \cite{Papadodimas:2015jra}. So a local low-energy infalling observer, described by a simple operator in the sense of \cite{Banerjee:2016mhh}, experiences the same physics irrespective of when she jumps into the wormhole. Only if she were a high-energy observer, i.e. a complicated operator scaling (at least) as $N$, she could distinguish the aforementioned states. The quantum correlations distinguishing the TFD and the TMD state can therefore be identified with $\frac{1}{N}$-corrections.

Let us highlight the fact that, via \eqref{eq:GQDRenyi}, the GQD is entirely determined by the second Rényi entropy. The bulk duals of Rényi entropies $S^{(n)}$ were derived in \cite{Dong:2016fnf} as the areas of cosmic branes. These cosmic branes have tension $T_n=\frac{n-1}{4nG_{\text{N}}}$ and back-react on the bulk geometry by creating conical defects $\Delta\phi=2\pi\frac{n-1}{n}$. In the limit where the conical defect vanishes, the corresponding cosmic brane becomes tension-less and reduces to the usual Ryu-Takayanagi minimal surface \cite{Dong:2016fnf}. In light of \eqref{eq:GQDRenyi}, this means that the bulk dual of the GQD can be related to a cosmic brane of tension $T_2=\frac{1}{8G_{\text{N}}}$ which creates a conical defect of $\Delta\phi=\pi$.

\section{\label{sec:Microstates}Geometric Quantum Discord Probing Black Hole Microstates}

\subsection{Microstate Overlaps and GQD}

We have seen above that quantum correlations are present as long as the partition function does not factorize, i.e. $Z(2\beta)\neq Z^2(\beta)$. Here we discuss the role of the black hole microstates $|\tfd\rangle_\alpha$ introduced in Sec.~\ref{sec:QDTFD} for this result \cite{Papadodimas:2015jra}. A particular set of phases $\{\alpha\}$ specifies the relation between the left and right boundary times and therefore also encodes how the left and right boundaries are glued to the bulk spacetime. Accordingly, we interpret computing the GQD as comparing one particular microstate $|\tfd\rangle_\alpha$ to the TMD state \eqref{eq:TMD}, the latter being the ensemble average of all microstates. In computing the GQD, the overlap between the microstate and the TMD state plays an important role: while the result \eqref{eq:GeometricMeasureForTFD} effectively looks like $1-\Tr(\rho_{\text{TMD}}^2)$, it is actually the overlap contribution $\Tr(\rho_{\text{TFD}_\alpha}\rho_{\text{TMD}})$ which is responsible for the relative sign in \eqref{eq:GeometricMeasureForTFD},
\begin{align}
    ||\rho_{\tfd_\alpha}-\rho_{\text{TMD}}||^2=1-2\Tr(\rho_{\text{TFD}_\alpha}\rho_{\text{TMD}})+\Tr(\rho_{\text{TMD}}^2)\,.
\end{align}
The fact that the phases do not influence the result for GQD is interpreted as an implicit averaging (i.e. self averaging) that is included in the definition of GQD by taking the trace.

The implicit averaging can be seen by computing the overlap explicitly for some general microstate $|\tfd\rangle_\beta$,
\begin{align}
    \Tr(\rho_{\tfd_\beta}\rho_{\text{TMD}})=\frac{1}{N}\sum_{\{\alpha\}}\,_\beta\langle\tfd|\tfd\rangle_\alpha\,_\alpha\langle\tfd|\tfd\rangle_\beta\,.\label{eq:overlap}
\end{align}
Each of the microstates is approximately, that is up to $\frac{1}{N}$-corrections, orthogonal to all other microstates, i.e. $\!\,_\beta\langle\tfd|\tfd\rangle_\alpha\simeq\delta_{\alpha\beta}$. However, precisely due to such $\frac{1}{N}$-corrections, averaging the square of the overlap does not simply equal a product of Kronecker symbols since evaluating the sum over the phases does not commute with approximating $\,_\beta\langle\tfd|\tfd\rangle_\alpha\simeq\delta_{\alpha\beta}$. This shows the non-factorizing property. In particular, before taking the states to be approximately orthogonal, the overlap is given by
\begin{align}
    \label{eqn:phase-avg}
    \,_\beta\langle\tfd|\tfd\rangle_\alpha=\sum_np_n^2e^{\i(\alpha_n-\beta_n)}\,.
\end{align}
Using this result in \eqref{eq:overlap} together with $\frac{1}{N}\sum_{\{\alpha\}}e^{\i(\alpha_n-\alpha_m)}=\delta_{nm}$ shows that the phases cancel exactly, with only $\sum_np_n^4$ remaining. The analogous argument is true when computing overlaps of a microstate with higher powers of the TMD state,
\begin{align}
    \Tr(\rho_{\tfd_\beta}\rho_{\text{TMD}}^n)&=\frac{1}{N^n}\sum_{\{\alpha\}_1,...,\{\alpha\}_n}\,_\beta\langle\tfd|\tfd\rangle_{\alpha_1}..._{\alpha_n}\langle\tfd|\tfd\rangle_\beta\notag\\
    &=\sum_mp_m^{2(n+1)}=\frac{Z\big((n+1)\beta\big)}{Z^{n+1}(\beta)}\,.\label{eq:HigherOverlaps}
\end{align}
This structure is highly reminiscent of the microstates discussed in \cite{Balasubramanian:2022gmo,Balasubramanian:2022lnw}, where the entropy of astrophysical black holes was studied. Since the collapse of a dust collection into a black hole is a unitary process, more realistic black holes also have a pure state description, at least immediately after the collapse and before decoherence sets in. Based on this, together with our proposal for generic pure states given in Sec.~\ref{sec:GQD}, we argue that our analysis extends to other pure states with a geometric dual. In particular, we can identify their results with ours in \eqref{eq:HigherOverlaps} in the limit $m_n\gg M$, $m_n$ and $M$ being the masses of a collapsing shell and a black hole, respectively. Therefore, the same arguments regarding the GQD hold, with a c-c state similar to TMD being the best classical approximation to the state dual to a black hole formed by collapse. Note however that in our case, we are considering an eternal black hole. By the above equation, we expect that the logic of \cite{Balasubramanian:2022gmo,Balasubramanian:2022lnw} can be applied to our scenario as well by interpreting a set of phases $\{\alpha\}_n$ as a collapsing shell of mass $m_n$. We can think of these $\alpha$ states as microstates of an eternal black hole contrary to the set up of \cite{Balasubramanian:2022gmo,Balasubramanian:2022lnw} where a time-reversal symmetry was assumed to construct the Hilbert space of microstates. Consequently, \eqref{eq:HigherOverlaps} can be interpreted as replica wormholes arising from state averaging \cite{Chakravarty:2020wdm,Freivogel:2021ivu,Goto:2021mbt}, contrary to conventional replica wormholes arising from ensemble averaging. This state averaging is manifest through the averaging over the phases as discussed below \eqref{eqn:phase-avg}. In fact, the normalization $\frac{1}{N}$ in the definition of the TMD state \eqref{eq:TMD} is very crucial here and emphasizes the fact that there are $N$ different sets of phases which can be interpreted as $N$ different microstates \cite{Verlinde:2020upt} leading to the thermodynamic entropy $S=\ln N$.

The higher overlaps \eqref{eq:HigherOverlaps}, which are interpreted as average of higher order spectral form factors in \cite{Verlinde:2021jwu}, also naturally appear when we generalize the (second power of the) Hilbert--Schmidt norm to the ($n$th power of the) $n$th Schatten norm,\footnote{Note that the Hilbert--Schmidt norm is related to the Schatten norm as the special case $n=2$.} defined for some matrix $X$ as
\begin{align}
    ||X||_{(n)}=\Big[\Tr(\sqrt{X^\dagger X\,}^{\,n})\Big]^{\frac{1}{n}}\,.
\end{align}
For a Hermitian matrix $X=X^\dagger$, the argument simplifies to $\sqrt{X^\dagger X}=X$. Evaluating the norm for $X=\rho_{\tfd_\alpha}-\rho_{\text{TMD}}=X^\dagger$ schematically results in
\begin{align}
    ||\rho_{\tfd_\alpha}-\rho_{\text{TMD}}||_{(n)}^n=1+\#_1\frac{Z(2\beta)}{Z^2(\beta)}+...+\#_n\frac{Z(n\beta)}{Z^n(\beta)}\,,
\end{align}
with $\#_k$ numerical factors ensuring that the norm vanishes if all of the partition functions factorize, i.e. $Z(k\beta)=Z^k(\beta)$ for $1\leq k\leq n$. This shows that all of these intermediate contributions have to be considered since all of them contribute to the presence of quantum correlations. Note also that analogous to above in \eqref{eq:GQDRenyi}, all of the ratios of partition functions can be expressed by higher Rényi entropies $S^{(n)}$ which, as mentioned before, can be calculated also on the gravity side by inserting cosmic branes of specific tension \cite{Dong:2016fnf}.

\subsection{Microstate Wormholes in Path Integral}
Our interpretation of non-vanishing GQD as the signature of wormholes arising from microstate averaging gets further strong support from the explicit structure of gravitational path integrals. As mentioned before, the presence of wormholes in the path integral of any quantum system is signaled by a non-exact symplectic form present in the path integral \cite{Verlinde:2021kgt}. The non-exactness argument was studied in the context of the TFD state in \cite{Nogueira:2021ngh,Banerjee:2022jnv}. In what follows, we will demonstrate the non-exactness of the symplectic form for a Hamiltonian which has the TFD states as its ground state. This will round up the direct one-to-one correspondence between the non-vanishing of GQD, the consequent non-factorization of the Hilbert space and the existence of wormhole microstates, as we elaborated in the previous sections. In particular, we provide an independent argument combining insights from \cite{Cottrell:2018ash} and \cite{Verlinde:2021kgt} from which it can be drawn that $Z(2\beta)\neq Z^2(\beta)$. We then compare our proposal for diagnosing non-factorization with the argument put forward in \cite{Verlinde:2021kgt} in view of practicality.

We start by making use of the construction of \cite{Cottrell:2018ash} to show that the Hamiltonian with the TFD state as its ground state possesses a non-exact symplectic form. For this, we summarize an independent argument for the non-factorization of the TFD state partition function using the methods developed in \cite{Cottrell:2018ash,Verlinde:2021kgt}. In \cite{Cottrell:2018ash}, the authors devise a mechanism to explicitly construct Hamiltonians with the TFD state as ground state by considering operators $d_k$ of the form
\begin{align}
    d_k=e^{-\frac{\beta}{4}(H_L^0+H_R^0)}\big(\mathcal{O}_{L,k}-\Theta\mathcal{O}_{R,l}^\dagger\Theta^{-1}\big)e^{\frac{\beta}{4}(H_L^0+H_R^0)}\,.
\end{align}
Here, $H_{L/R}^0$ and $\mathcal{O}_{L/R,k}$ are the Hamiltonians and any operators of the left and right theories, respectively, and $\Theta$ is the same anti-unitary operator that appeared in \eqref{eq:TFD}. Since $d_k|\tfd\rangle=0$ by construction, the Hamiltonian
\begin{align}
    H=c_kd_k^\dagger d_k\,,
\end{align}
has the TFD state as its ground state, where $c_k$ are positive constants. Since the exponentials of the Hamiltonians are hard to compute for strongly coupled systems, a simpler version of the above operators,
\begin{align}
    d_k=\mathcal{O}_{L,k}-\Theta\mathcal{O}_{R,k}^\dagger\Theta^{-1}\,,
\end{align}
is used to define
\begin{align}
    H=H_L^0+H_R^0+c_k^\prime d_k^\dagger d_k\,,
\end{align}
where $c_k^\prime$ are again positive constants, but have to be chosen appropriately. Assuming that the eigenvalue thermalization hypothesis holds, it was shown that also this simpler Hamiltonian has the TFD state as its ground state \cite{Cottrell:2018ash}. In the simple case of two harmonic oscillators, the two Hamiltonians constructed above coincide, provided that $c_k^\prime$ are chosen appropriately.

Explicitly, the Hamiltonian for two harmonic oscillators is given by \cite{Cottrell:2018ash}
\begin{align}
    H&=\frac{1+c}{2}\big(p_L^2+p_R^2\big)+\frac{(1+c)\omega^2}{2}\big(q_L^2+q_R^2\big)+cp_Lp_R-c\omega^2q_Lq_R\,,
\end{align}
where $c=\frac{1}{2}\csch^2\frac{\beta\omega}{4}$. Following the diagonalization of this Hamiltonian into variables $p_\pm,q_\pm$, we introduce creation and annihilation operator to write the Hamiltonian as
\begin{align}
    H=\epsilon_+a_+^\dagger a_++\epsilon_-a_-^\dagger a_-\,,
\end{align}
where $\epsilon_+=\frac{1}{\omega}\coth\frac{\beta\omega}{4}$ and $\epsilon_-=\frac{1}{\omega}\tanh\frac{\beta\omega}{4}$. To manifestly write the Hamiltonian in the form discussed in \cite{Verlinde:2021kgt}, we introduce operators $J=a_k^\dagger a_k$ and $X_I=\frac{1}{2}a_k^\dagger\sigma_I^{kl}a_l$ such that $J^2=\sum_{I=1}^3X_I^2$ to find the Hamiltonian
\begin{align}
    H=\gamma J+\epsilon X_3\,,
\end{align}
where $\gamma=\frac{1}{\omega}\coth\frac{\beta\omega}{2}$ and $\epsilon=\frac{1}{\omega}\csch\frac{\beta\omega}{2}$. Since this Hamiltonian with the TFD state as its ground state is of the class of Hamiltonians discussed in \cite{Verlinde:2021kgt}, the Hamiltonian is associated with a non-exact symplectic form when the system is restricted to a quantizable orbit, $J=j$ with $2j\in\mathds{N}$. Explicitly, the symplectic form in this case can be understood as the volume form of the two-sphere, which is non-exact. The non-exactness implies that the corresponding partition function does not factorize but includes replica wormholes \cite{Verlinde:2021kgt}. This constitutes an independent argument for the non-factorization of the TFD Hilbert space and agrees with our expectations from computing the GQD for the TFD state.

Let us now compare our method for diagnosing non-factorization via GQD with that put forward in \cite{Verlinde:2021kgt} in view of ease of implementation. To prove that a form $\omega$ is exact, one has to show that there exists a closed surface $S$ such that the integral $\int_S\omega$ vanishes. Finding an appropriate surface $S$ can be complicated in general, especially when the phase space is large. Therefore, diagnosing non-factorization by the non-exactness of the symplectic form might be hard in practice given a particular system. In these cases, our criterion of non-zero GQD might be easier to compute as explained in Sec.~\ref{sec:GQD} for generic pure states.

\section{\label{sec:Conclusions}Conclusion}

In this paper, we propose a novel procedure to diagnose boundary non-factorization based on a geometric measure of quantum correlations called geometric quantum discord \eqref{eq:GeometricQuantumDiscord}. By computing the GQD explicitly for generic pure states and for the TFD state as a holographic example, we show that the non-vanishing of the GQD signals non-factorization in terms of a non-factorizing modular or thermal partition function, respectively. Our analysis of the TFD state relates GQD to wormholes, the latter implying non-factorization from a bulk perspective. Moreover, we identify the TMD state as the state minimizing the GQD for the TFD state, thus describing the \emph{closest classical approximation to the TFD state}. This non-trivial statement is independent of any holographic interpretation and shows that the minimization involved in GQD necessarily leads to classical states, as we find also for generic pure states apart from the TFD state. This new quantitative analysis of the correlations within the TFD state provides a novel path to further understanding the factorization puzzle. Furthermore, we have proven the equality of two geometric measures of quantum correlations, the GQD $Q^{(2)}(A\!:\!B)$ \eqref{eq:GeometricQuantumDiscord} and the measure $\bar{Q}^{(2)}(A\!:\!B)$ introduced in \cite{Guo:2020rwj}, for the case of pure states. Additionally, we show that this non-vanishing of GQD is a direct consequence of state-averaged replica wormholes which show up in the gravitational path integral in the form of a non-exact symplectic form. We demonstrated this explicitly using a quantum mechanical Hamiltonian which yields a TFD state as its ground state. Finally, we encounter similarities between higher order overlaps of wormhole microstates considered in \cite{Balasubramanian:2022gmo,Balasubramanian:2022lnw} and generalizations of the Hilbert-Schmidt norm, known as Schatten norms. We expect these norms to also have an information-theoretic interpretation akin to that of GQD, but we are not aware of any explicit results in this regard.

Our results open several avenues for further research. First, it would be interesting to investigate GQD in the context of field theory, particularly for two-dimensional CFTs for which we have argued that the replica trick provides a powerful tool to access it in view of our result \eqref{eq:GQDRenyi}. The generators of the Virasoro algebra represent a canonical measurement basis. This is promising in view of developing a consistent notion of projective measurements \cite{Antonini:2022lmg,Antonini:2023aza}, a necessary step towards generalizing quantum discord itself to field theories. In a more holographic context, the bulk computation of GQD in terms of cosmic branes is an exciting task to pursue.

Second, the TMD state \eqref{eq:TMD} considered here was referred to as \enquote{old TMD} state in \cite{Verlinde:2021jwu} and it was argued that it describes the plateau region of the spectral form factor. In the same work, a \enquote{new TMD} state was defined in which not all quantum correlations have yet been lost to decoherence, and an argument for its description of the dip region was given. We expect the computation of GQD for the new TMD state to be a promising step towards a better understanding of the dynamics of decoherence, and we leave it for future studies. In a similar vein, the identification of explicit quantum information-theoretic measures which capture the structure of the higher overlaps obtained from Schatten norms is an interesting task to follow. Progress in this direction would help to further expand the role of quantum information theory within the holographic dictionary, and potentially provide a finer classification of the correlations giving rise to the factorization puzzle.

Finally, it is intriguing to investigate the GQD for mixed states, where entanglement entropy falls short. For mixed states, the minimization will in general not lead to a c-c state, as we have discussed for the isotropic state of a two-spin system. Consequently, the two measures $Q^{(2)}(A:B)$ and $\bar{Q}^{(2)}(A:B)$ will differ for mixed states. Though mixed states are not so frequent within holography, one way to generate them is by performing a double bipartition of a pure state. First, separate the full system into $A$ and $B$, and subsequently separate $A$ into $A_1$ and $A_2$. For such a setting, it will be interesting to compare GQD with entanglement negativity, the latter one being computed for the BTZ black hole e.g. in \cite{Chaturvedi:2016rcn}. We leave an investigation of $Q^{(2)}(A_1:A_2)$ in this direction for future work.

\acknowledgments

We thank Chris Akers, Mohsen Alishahiha, Giuseppe Di Giulio, Johanna Erdmenger, Haye Hinrichsen, Nima Lashkari, René Meyer, Christian Northe and Anna-Lena Weigel for stimulating discussions. We also thank Mohsen Alishahiha, Johanna Erdmenger and Haye Hinrichsen, in particular, for helpful comments on the draft. We acknowledge support by the Deutsche Forschungsgemeinschaft (DFG, German Research Foundation) under Germany's Excellence Strategy through the Würzburg-Dresden Cluster of Excellence on Complexity and Topology in Quantum Matter - ct.qmat (EXC 2147, project-id 390858490). We further acknowledge the support by the Deutscher Akademischer Austauschdienst (DAAD, German Academic Exchange Service) through the funding programme, \enquote{Research Grants - Doctoral Programmes in Germany, 2021/22 (57552340)}. This research was also supported in part by Perimeter Institute for Theoretical Physics. Research at Perimeter Institute is supported by the Government of Canada through the Department of Innovation, Science and Economic Development and by the Province of Ontario through the Ministry of Research, Innovation and Science.

\bibliographystyle{JHEP}
\bibliography{DiscordBibliography.bib}

\end{document}